# Estimation of mitral valve hinge point coordinates - deep neural net for echocardiogram segmentation

Christian Schmidt

Westfälische Hochschule –
University of Applied Sciences
Medical Engineering Laboratory

Neidenburger Str. 43
45897 Gelsenkirchen, Germany

christian.schmidt@w-hs.de

Heinrich Martin Overhoff

Westfälische Hochschule –
University of Applied Sciences
Medical Engineering Laboratory

Neidenburger Str. 43
45897 Gelsenkirchen, Germany

heinrich-martin.overhoff@w-hs.de

## ABSTRACT

Cardiac image segmentation is a powerful tool in regard to diagnostics and treatment of cardiovascular diseases. Purely feature-based detection of anatomical structures like the mitral valve is a laborious task due to specifically required feature engineering and is especially challenging in echocardiograms, because of their inherently low contrast and blurry boundaries between some anatomical structures. With the publication of further annotated medical datasets and the increase in GPU processing power, deep learning-based methods in medical image segmentation became more feasible in the past years. We propose a fully automatic detection method for mitral valve hinge points, which uses a U-Net based deep neural net to segment cardiac chambers in echocardiograms in a first step, and subsequently extracts the mitral valve hinge points from the resulting segmentations in a second step. Results measured with this automatic detection method were compared to reference coordinate values, which with median absolute hinge point coordinate errors of 1.35 mm for the *x*- (15-85 percentile range: [0.3 mm; 3.15 mm]) and 0.75 mm for the *y*- coordinate (15-85 percentile range: [0.15 mm; 1.88 mm]).

## 1. INTRODUCTION

According to the World Health Organization (WHO), 17.9 million people died from cardiovascular diseases (CVDs) in 2019, which is 32% of all global deaths. Advances in medical imaging significantly improved the process of diagnostics and treatment of CVDs over the past years, with cardiac image segmentation playing an important role.

Cardiac image segmentation is the process of partitioning an image by assigning a label to each pixel of the image in such a way, that pixels of a certain anatomical structure share the same label. Anatomical and functional parameters such as left ventricle (LV) volume, left atrium (LA) volume, ejection fraction and mitral valve (MV) dimensions can be determined using segmented images.

Direct segmentation of the two mitral valve leaflets (MVLs) with purely feature-based algorithms often fails, because of low contrast in echocardiograms or lack of visualization of both MVLs at the same time, which is common in clinical settings. Therefore, we propose to assess MV hinge point coordinates by using deep learning (DL) segmentations of the LV and LA. In 2019 Leclerc et al. [Lec19] published the Cardiac Acquisitions for Multi-structure Ultrasound Segmentation (CAMUS) dataset in conjunction with an image segmentation challenge https://www.creatis.insa-lyon.fr/Challenge/camus/.
This is to our knowledge the first large-scale, publicly available transthoracic echocardiography (TTE) dataset, which includes ground truth segmentations of the LV and LA.

In this paper, we use the CAMUS dataset (Fig. 1, Fig. 2) to segment the LV and LA in apical four (a4c) and two-chamber views (a2c) in a first step, and in a second step extract the mitral valve diameter and hinge point coordinates from the resulting segmentation.

To the best of our knowledge, this is the first attempt of DL-based mitral valve measurement in transthoracic four- and two-chamber view echocardiograms.

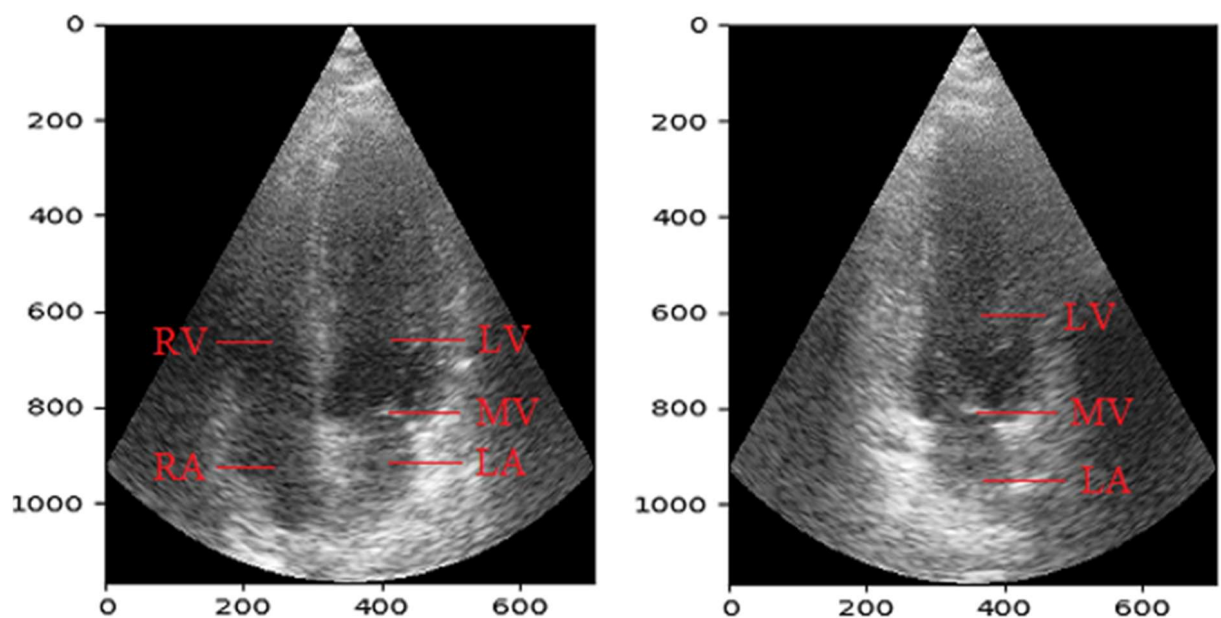

**Figure 1. Exemplary labelled TTE of the CAMUS dataset in a4c (left) and a2c view (right), showing the ventricles (LV and RV), atria (LA and RA) and the mitral valve (MV).**

## 2. RELATED WORKS

Conventional machine learning techniques, e.g., active shape and atlas-based models [Okt14; Tav13], showed good performance in cardiac image segmentation, but rely on user-based, manual feature extraction.

In recent years, with the increased availability of high-performance GPUs and more access to image training data, deep learning (DL) models, which automatically learn features from image data, have outperformed conventional machine learning techniques.

These DL segmentation approaches mainly consist of encoder-decoder convolutional neural networks (CNN), in particular fully convolutional networks [Lon15] and the U-Net architecture [Ron15], using ResNet [HeK16], Inception [Sze15] or VGG [Sim15] as popular encoder backbones, as these have performed best in the field of medical image segmentation.

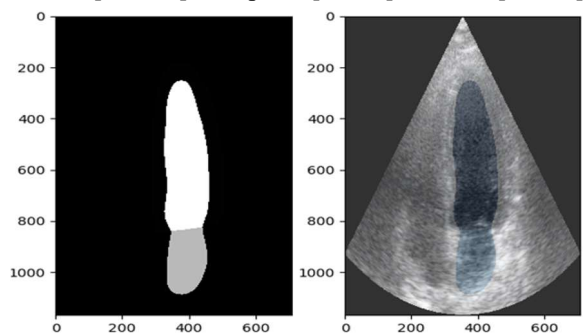

**Figure 2. Ground truth annotation of the LV and LA in a4c image (left), overlay of the annotation on the original echocardiogram (right).**

TTE is the most commonly performed imaging examination in cardiology, due to the fact that it is non-invasive, has low cost and high accessibility, yet up to 80% of annual publications between 2016 and 2019 [Che20] on DL-based cardiac image segmentation worked with magnetic resonance imaging (MRI) data [YuL17; Wol17], mainly because of larger dataset availability. Computer tomography (CT) [Zre15; Ton17] and echocardiography [Zha18; Smi17], despite their clinical importance, only played a subordinate role due to the lack of annotated datasets.

As for MV measurement in particular, in clinical practice, MV dimensions are either manually obtained by a user manually selecting points on frozen frames throughout the cardiac cycle [Gar15; Dwi14] or by semi-automatic segmentation. E.g., [Pou12] proposes an MV morphometry method, which requires user initialized selection of a region of interest and anatomical landmarks followed by feature-based contour segmentation.

Further feature-based (semi-)automatic methods for MV assessment often require vendor-specific software for analysis. In addition, they have high computational run times and have only been assessed in single-center studies with small patient numbers and little variety in MV conditions [Nol19].

## 3. PROPOSED SOLUTION

### 3.1 Dataset

The CAMUS dataset includes 450 annotated patient sub-datasets consisting of TTE apical four- and two-chamber views. Each patient sub-dataset consists of one cardiac cycle per view, but ground truth segmentations are only provided at the image frames at end-diastole (ED) and end-systole (ES). Image sizes vary, in a range between 400 x 700 pixels and 700 x 1000 pixels, with a spatial grid resolution of 0.3 mm along the *x*- and 0.15 mm along the *y*-axis. The dataset includes images of various acquisition settings, e.g., resolution, image contrast and transducer angle. Furthermore, some images are not properly centered, therefore certain anatomical structures (ventricles, atria) are potentially not fully visible on them. No further data selection or preprocessing has been performed. This results in a heterogeneous dataset, which is a realistic representation of data acquired in clinical practice.

Manual delineation of the LV, LA, and epicardium was performed by a cardiologist using a defined segmentation protocol. In particular, the LV delineation contour was to be terminated at the points where the MVLs are hinging. Epicardium annotation is not relevant to mitral valve measurement and is therefore not considered in this work.

### 3.2 Training the segmentation model

We introduce a two–step method for estimating MV hinge point coordinates. This method uses a deep learning algorithm for segmentation of the left ventricle and left atrium and subsequent feature-based image processing for the estimation of MV hinge point coordinates.

*Step 1: deep learning segmentation algorithm*

Since [Lec19] demonstrated that the U-Net architecture showed slightly better segmentation accuracy on the CAMUS dataset than more sophisticated encoder-decoder networks, a U-Net with the VGG16 backbone was used to train our model. Model training was implemented in Python version 3.7.7 using Tensorflow 2.0.0 in conjunction with the Keras API. The Adam optimizer [Kin14], a learning rate of $10^{-3}$ and the categorical cross-entropy loss function were used for training. No data augmentation was performed on the dataset.

450 patient sub-datasets were divided into three groups, 350 for training, 50 for validation, and 50 for testing, a roughly 80%/10%/10% split. Model training and validation were performed on an NVIDIA GeForce RTX 2060 and ran for 50 epochs, after which the validation accuracy stagnated or dropped, and training was terminated to avoid overfitting (Fig. 3). As a result of this first step, image pixels are assigned

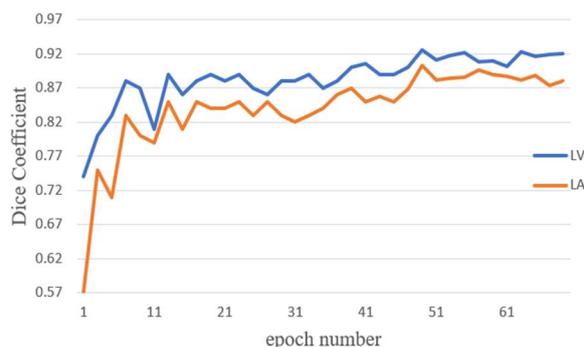

**Figure 3. Dice coefficient of the left ventricle and left atrium for the validation dataset, training was stopped after 50 epochs.**

to the left ventricle (LV), left atrium (LA), and background.

### 3.3 Extraction of mitral valve hinge points

*Step 2: feature-based hinge point extraction*

A purely feature-based algorithm for mitral valve detection, which uses e.g., thresholding, edge-detection, or histogram methods, is likely to perform insufficiently in regions of low pixel grayscale gradients, and thus does not detect both MVLs reliably.

In numerous clinical cases, anatomical structures are not clearly visible on echocardiograms, due to low image contrast and blurry boundaries between anatomical structures.

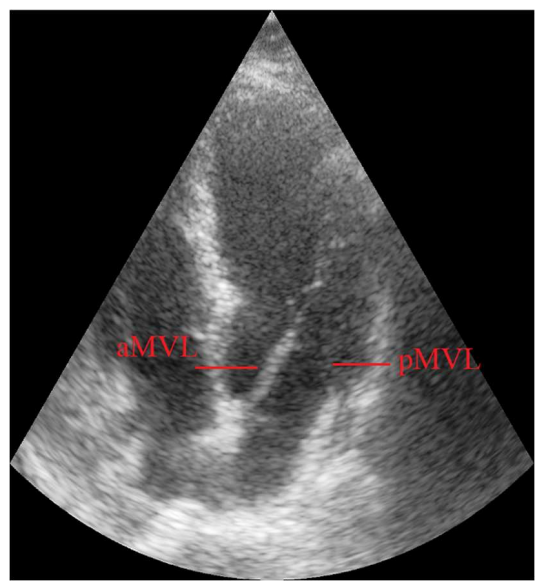

**Figure 4. Apical four-chamber view (a4c) at end diastole (ED). The intersecting line between image plane and anterior mitral valve leaflet (aMVL) is clearly visible, whereas the posterior leaflet (pMVL) is barely visible.**

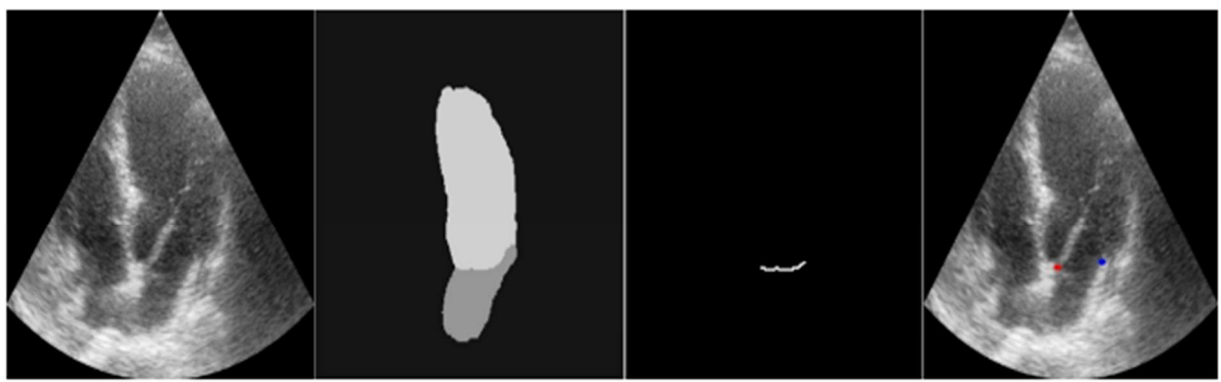

a) original echocardiogram   b) LA, LV segmentation   c) extracted contact line   d) hinge points overlay

**Figure 5. Overview of the proposed method: The original echocardiogram (a) is first segmented, using the CNN. The resulting contact line (c) between the segments of left ventricle and left atrium (b) is then used to extract the mitral valve hinge points (d).**

In particular, MVLs are hardly distinguishable from the background in many cases (Fig. 4).

Therefore, we use the DL-generated segmentations of a4c- and a2c echocardiograms (see section 3.2) to estimate MV hinge point coordinates. Figure 5 gives an overview of the individual steps in our proposed method.

According to the contouring protocol in [Lec19], the LV contour was to be terminated in the MV plane, at the MV hinge points. Using the resulting contact line between LV and LA segmentation (Fig. 5c), we define the anteriormost point of the contact line as the anterior mitral valve leaflet (aMVL) and the posteriormost point of the contact line as the posterior mitral valve leaflet (pMVL).

This second step results in *x*- and *y*-coordinates for the aMVL and pMVL.

### 3.4 Evaluation Metrics

The segmentation CNN in combination with the following MV hinge point extraction, is to be considered as a measurement tool for the pixel-coordinates of the MV hinge points. In this method, each measurement $\hat{z}$ is determined by

$$\hat{z} = z + \Delta\hat{z} = z + \left(\Delta\hat{z}^{(\text{bias})} + e^{(\hat{z})}\right)$$

where $z$ is the true value, $\Delta\hat{z}^{(\text{bias})}$ the systematic error and $e^{(\hat{z})}$ the random error (DIN 1319-1/ISO 11843-1). Typically, a normal distribution of errors $\Delta\hat{z}$ is assumed and characterized by its mean $\mu$ and standard deviation $\sigma$. To assess the normality of our error distributions, we performed Shapiro-Wilk tests for $\Delta\hat{z}$ data series of each subgroup ($\Delta\hat{x}_{\text{aMVL}}$, $\Delta\hat{y}_{\text{aMVL}}$, $\Delta\hat{x}_{\text{pMVL}}$, $\Delta\hat{y}_{\text{pMVL}}$).

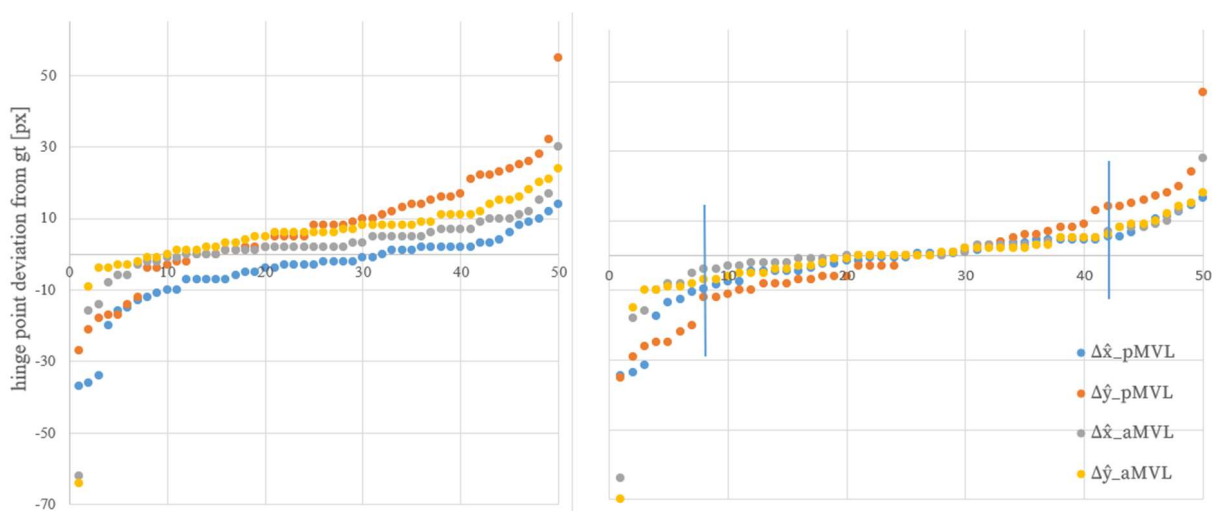

**Figure 6. Sorted hinge point coordinate errors [px] before (left) and after (right) calibration of the systematic error $\Delta\hat{z}^{(\text{bias})} = \left\{\Delta\hat{x}_{\text{pMVL}}^{(\text{bias})}, \Delta\hat{y}_{\text{pMVL}}^{(\text{bias})}, \Delta\hat{x}_{\text{aMVL}}^{(\text{bias})}, \Delta\hat{y}_{\text{aMVL}}^{(\text{bias})}\right\}$. The blue vertical lines in the right diagram show the 15th and 85th percentiles, which are evaluated in the results section.**

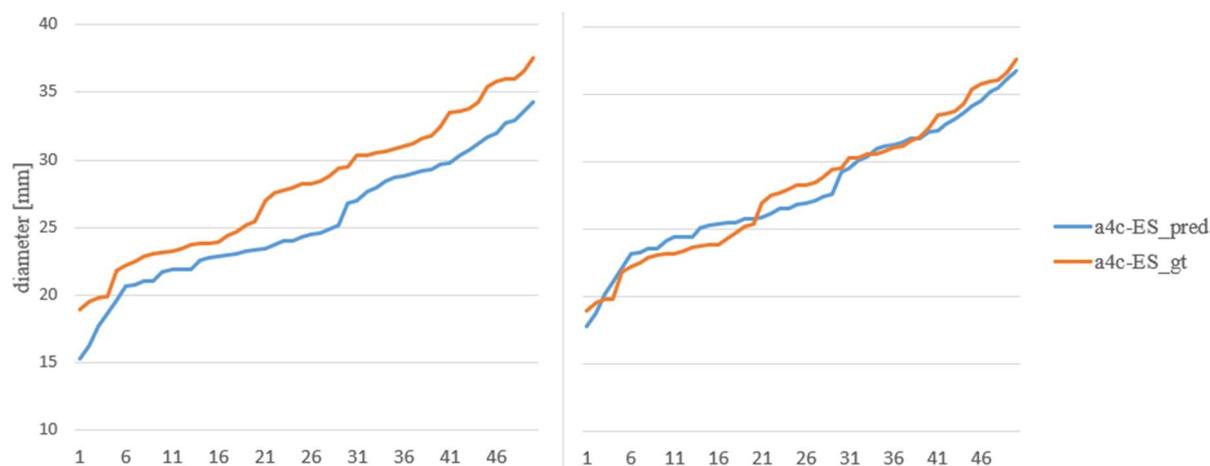

**Figure 7. Sorted ground truth and predicted MV diameters (mm) before (left) and after (right) calibration of the systematic error $\Delta\hat{z}^{(\text{bias})}$. The predicted MV diameters were systematically underestimated by 13.8%.**

These tests resulted in *p*-values of $p < 0.05$ for all but one data series, thus the assumption of normal distribution is rejected.

Therefore, we characterize the distributions by their 15-, 50- (median), and 85-percentiles instead, as equivalents to $\mu - \sigma$, $\mu$ and $\mu + \sigma$.

To account for systematic errors $\Delta\hat{z}^{(\text{bias})}$ and subsequently only evaluate random errors $e^{(\hat{z})}$, median deviations $\Delta\hat{z}^{(\text{bias})}$ ($\Delta\hat{x}_{\text{aMVL}}^{(\text{bias})}$, $\Delta\hat{y}_{\text{aMVL}}^{(\text{bias})}$, $\Delta\hat{x}_{\text{pMVL}}^{(\text{bias})}$, $\Delta\hat{y}_{\text{pMVL}}^{(\text{bias})}$) are subtracted from measured values $\hat{z}$ (calibration). We then evaluate the 15-85 percentile range of calibrated values. (Fig. 6).Calibrated, random *x*- and *y*- coordinate errors of the aMVL and the pMVL ($e_{\text{aMVL}}^{\hat{x}}, e_{\text{aMVL}}^{\hat{y}}, e_{\text{pMVL}}^{\hat{x}}, e_{\text{pMVL}}^{\hat{y}}$) will be evaluated individually.

In addidition, the segmentation accuracy of the CNN as well as the resulting MV diameters will be evaluated.

## 4. RESULTS
### 4.1 Chamber segmentation accuracy

To evaluate the segmentation accuracy of our network, we use the Dice Coefficient (*D*).

|  | $D_{\text{LV\_ED}}$ | $D_{\text{LV\_ES}}$ |
|---|---|---|
| proposed | 0.931 | 0.915 |
| [Lec19] | 0.939 | 0.916 |

**Table 1. Dice Coefficients of LV U-Net segmentation.**

The combined Dice Coefficient of all LVs (at ED and ES) is $D_{\text{LV}} = 0.923$, with segmentations at ED performing slightly better than at ES ($D_{\text{LV\_ED}} = 0.931$, $D_{\text{LV\_ES}} = 0.915$). Unlike the procedure described in [Lec19], we did not perform any post-processing (e.g. connected component analysis) on the segmentation result, yet are still in line with their best U-Net

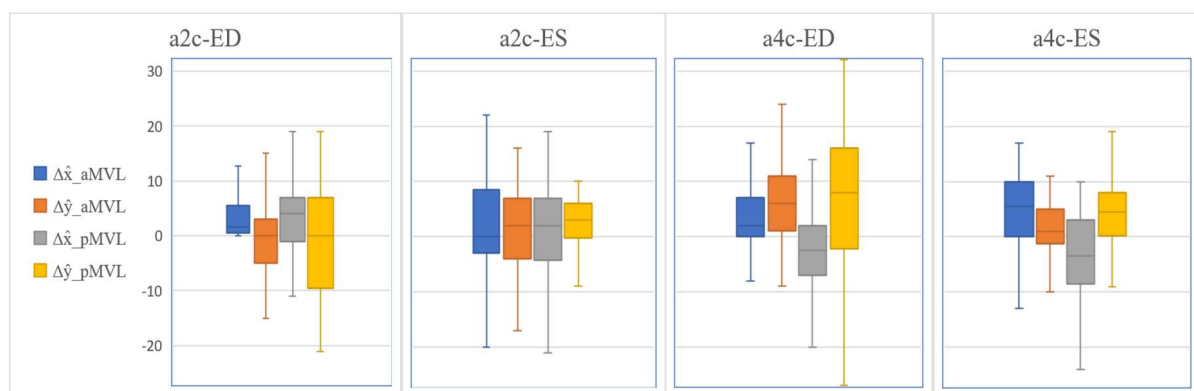

**Figure 8. Boxplot diagrams of hinge point coordinate errors $\Delta\hat{x}_{\text{aMVL}}, \Delta\hat{y}_{\text{aMVL}}, \Delta\hat{x}_{\text{pMVL}}, \Delta\hat{y}_{\text{pMVL}}$ [px] of a4c, a2c images at ES and ED before calibration.**

segmentation accuracy ($D_{LV\_ED}$ = 0.939,　$D_{LV\_ES}$ = 0.916).

## 4.2 Mitral valve annulus diameter

Our experimental measurements of the MV annulus diameters (Table 1) lie well within the range of empiric MV measurement data. E.g., in [Dwi14], 5-95 percentile ranges for the mitral annulus diameter of 22-38 mm are stated for women and 25-41 mm for men.

While median diameter estimations for a2c images and a4c images (Table 2) at ED conformed well with ground truth values, a significant systematic estimation bias was observed in a4c images at ES. Here, the median diameter was underestimated by 13.8% (Fig. 7). This systematic error can also be seen in the individual hinge point coordinates in section 4.3.

|        | predicted | ground truth |
|--------|-----------|--------------|
| a4c-ED | 27.9 mm   | 28.8 mm      |
| a4c-ES | 24.4 mm   | 28.3 mm      |
| a2c-ED | 31.3 mm   | 31.7 mm      |
| a2c-ES | 26.3 mm   | 26.1 mm      |

**Table 2. Predicted and ground truth median MV annulus diameter values a2c and a4c views at ES and ED.**

## 4.3 MV hinge point coordinates

Evaluation of individual hinge point coordinate errors (Fig. 8) shows further systematic estimation errors $\Delta\hat{z}^{(bias)}$. Since both the anterior and posterior hinge points are estimated too far inwardly (medial) in a4c view at ES, the corresponding underestimation in diameters (see section 4.2) is explained.

Figure 9 displays the individual systematic errors $\Delta\hat{z}^{(bias)}$ for the aMVL- and pMVL hinge point of each view type. The coordinate estimation is, on average, biased towards the bottom (0.5 mm) and the right (0.3 mm). The estimation accuracy in terms of absolute coordinate error distance in mm was much lower for the *x*- compared to the *y*-coordinate, as can be seen in Table 3.

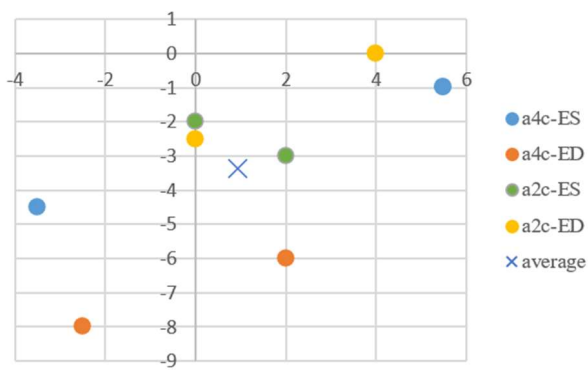

**Figure 9. Systematic errors $\Delta\hat{z}^{(bias)}$ [px] of each individual subgroup (a4c-ED, a4c-ES, a2c-ED, a2c-ES). On average (marked by blue x) hinge points are estimated about 0.3 mm too far posterior (in the image: right) and about 0.5 mm too far cranial (in the image: down).**

This is almost fully explained by the spatial resolution of the images, which is 0.3 mm along the *x*- and 0.15 mm along the *y*- axis, as described in section 3.1.

This results in absolute median coordinate errors of 1.35 mm for all *x*-coordinates and 0.75 mm for all *y*-coordinates. When comparing the median of absolute error distances median($|e^{(\hat{z})}|$) of the different views (a4c-ED, a4c-ES, a2c-ED, a2c-ES), estimation accuracy was approximately equal in the four subgroups.

## 4.4 Impact of off-center images

Looking at the correlation plot (Fig. 10) between predicted and ground truth $x_{aMVL}$, $x_{pMVL}$ coordinates of the MV hinge points in a4c views, a subdivision of data points into two groups can be observed. We suspect this is likely due inaccurate centering of the displayed portion of the a4c view. Since most of the misaligned, atypical a4c images are heavily centered on the LV (Fig. 11), the LA is not properly displayed, which leads to lower segmentation accuracy and thus higher estimation errors of the MV hinge point coordinates. No similar phenomenon of data subdivision was observed in hinge point coordinates in a2c view.

|        | $e^{\hat{x}}_{aMVL}$ |              | $e^{\hat{y}}_{aMVL}$ |              | $e^{\hat{x}}_{pMVL}$ |              | $e^{\hat{y}}_{pMVL}$ |              |
|--------|----------------------|--------------|----------------------|--------------|----------------------|--------------|----------------------|--------------|
|        | signed               | median(abs)  | signed               | median(abs)  | signed               | median(abs)  | signed               | median(abs)  |
| a4c-ED | [−1.2; 1.5]          | 1.35         | [−0.9; 0.75]         | 1.2          | [−2.2; 1.35]         | 0.9          | [−1.65; 1.8]         | 0.75         |
| a4c-ES | [−2.3; 1.95]         | 1.65         | [−0.75; 0.9]         | 0.53         | [−2.55; 2.3]         | 1.8          | [−1.8; 0.83]         | 0.53         |
| a2c-ED | [−1.5; 1.2]          | 1.35         | [−1.8; 1.35]         | 0.83         | [−2.1; 2.1]          | 1.2          | [−1.88; 0.8]         | 1.05         |
| a2c-ES | [−1.8; 2.8]          | 1.65         | [−1.35; 0.9]         | 0.45         | [−2.6; 2.1]          | 1.2          | [−1.2; 0.75]         | 0.75         |

**Table 3. Results of hinge point estimations [mm]. In addition to the signed 15-85 percentile ranges of $e^{(\hat{z})}$ ($e^{\hat{x}}_{aMVL}$, $e^{\hat{y}}_{aMVL}$, $e^{\hat{x}}_{pMVL}$, $e^{\hat{y}}_{pMVL}$), the median of absolute error distances median($|e^{(\hat{z})}|$) is stated.**

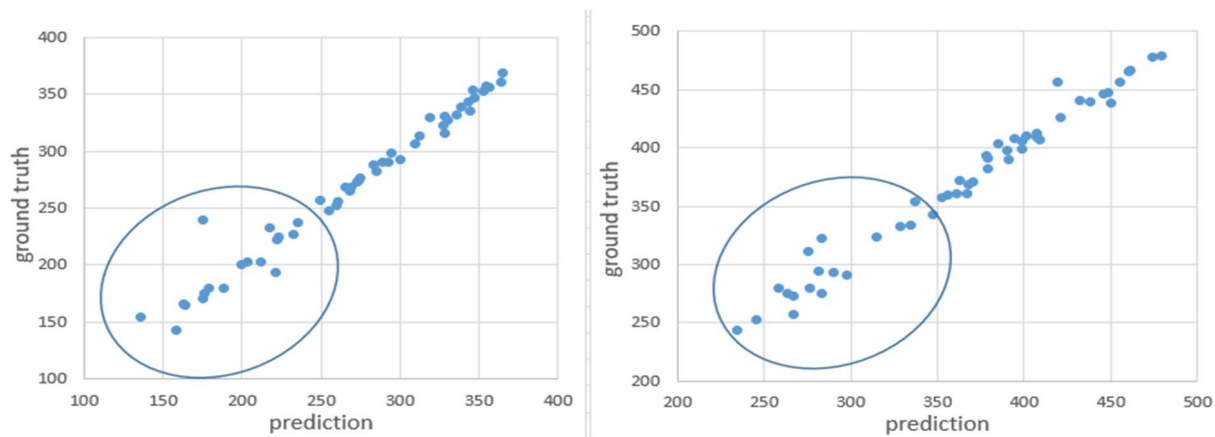

**Figure 10. Correlation plot of ground truth and predicted $x_{\text{aMVL}}$ (left) and $x_{\text{pMVL}}$ (right) of a4c images at ED. The circled subgroup of hinge points is positioned far more anterior than the rest of the datapoints and shows higher deviations from the ground truth.**

## 5. CONCLUSION

We demonstrated a two–step method for estimating MV hinge point coordinates using deep learning segmentations of the left ventricle and left atrium and subsequent feature based image processing. With 15-85 percentile ranges of random coordinate estimation errors $e^{(\hat{z})}$ between [−0.9 mm; 0.75 mm] and [−2.55 mm; 2.3 mm] and absolute median coordinate errors of 1.35 mm and 0.75 mm respectively, the resulting estimations are satisfactory, but further improvements can be made.

If the LV and LA can be adequately segmented by the neural net, the resulting segmentation mask can be used to reliably determine the MV hinge point coordinates.

We used the CAMUS dataset in this work, which is quite heterogeneous and, as such, close to clinical practice, as described above. This is beneficial for the generalizability of the network. On the other hand, the heterogeneity (e.g., low-quality images, edge cases described in section 4.4) is detrimental to estimation accuracy. Depending on the use case, adjustments to the training data set can be made. If generalizability is the highest priority, further low-quality and off-centered images should be added to adequately represent them in the training data. Otherwise, if estimation accuracy is the priority, low-quality images can be removed from the dataset with instructions to the physician to record more appropriate images.

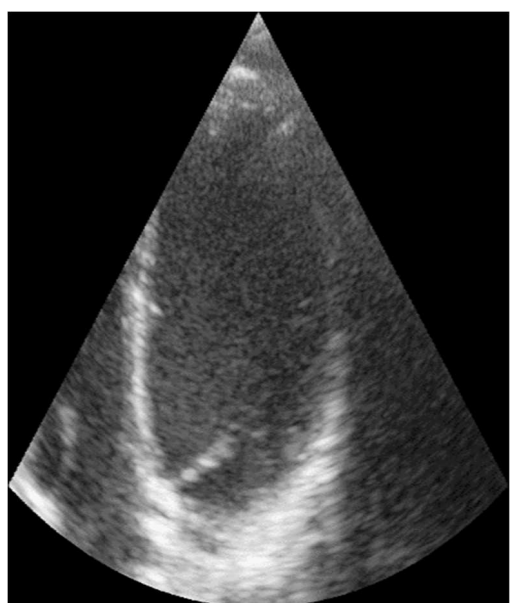

**Figure 11. Example of inaccurately centered image of the four-chamber view, only the LV and part of the RV are visible.**